\documentclass[aps,prb,twocolumn,groupedaddress,preprintnumbers,amsmath,assymb]{revtex4}

\usepackage[english]{babel}
\usepackage{graphicx}
\usepackage{subfigure}
\usepackage{dcolumn}
\usepackage{bm}
\usepackage{psfrag}

\begin{document}

\title{Exact ground state Monte Carlo method for Bosons without importance sampling}

\author{M. Rossi}
\email[]{maurizio.rossi@unimi.it}
\affiliation{Dipartimento di Fisica, Universit\`a degli Studi di Milano,
             via Celoria 16, 20133 Milano, Italy}
\author{M. Nava}
\affiliation{Dipartimento di Fisica, Universit\`a degli Studi di Milano,
             via Celoria 16, 20133 Milano, Italy}
\author{L. Reatto}
\affiliation{Dipartimento di Fisica, Universit\`a degli Studi di Milano,
             via Celoria 16, 20133 Milano, Italy}
\author{D.E. Galli}
\affiliation{Dipartimento di Fisica, Universit\`a degli Studi di Milano,
             via Celoria 16, 20133 Milano, Italy}
\date{\today}

\begin{abstract}
Generally ``exact'' Quantum Monte Carlo computations for the ground state of many
Bosons make use of importance sampling. 
The importance sampling is based, either on a guiding function or on an initial 
variational wave function.
Here we investigate the need of importance sampling in the case of Path Integral 
Ground State (PIGS) Monte Carlo.
PIGS is based on a discrete imaginary time evolution of an initial wave function
with a non zero overlap with the ground state, that gives rise to a discrete path 
which is sampled via a Metropolis like algorithm.
In principle the exact ground state is reached in the limit of an infinite imaginary
time evolution, but actual computations are based on finite time evolutions and the 
question is whether such computations give unbiased exact results.
We have studied bulk liquid and solid $^4$He with PIGS by considering as initial wave 
function a constant, i.e. the ground state of an ideal Bose gas.
This implies that the evolution toward the ground state is driven only by the 
imaginary time propagator, i.e. there is no importance sampling.
For both the phases we obtain results converging to those obtained by considering
the best available variational wave function (the Shadow wave function) as initial
wave function.
Moreover we obtain the same results even by considering wave functions with the
wrong correlations, for instance a wave function of a strongly localized Einstein 
crystal for the liquid phase.
This convergence is true not only for diagonal properties such as the energy, the radial 
distribution function and the static structure factor, but also for off--diagonal 
ones, such as the one--body density matrix.
This robustness of PIGS can be traced back to the fact that the chosen initial wave 
function acts only at the beginning of the path without affecting the imaginary time
propagator.
From this analysis we conclude that zero temperature PIGS calculations can be as 
unbiased as those of finite temperature Path Integral Monte Carlo.
On the other hand, a judicious choice of the initial wave function greatly improves
the rate of convergence to the exact results.
\end{abstract}

\maketitle

\section{INTRODUCTION}
\label{sec:intro}

Among the available methods to investigate the properties of strongly interacting many--body 
quantum systems, Quantum Monte Carlo (QMC) ones hold a relevant position since some of the
QMC methods can provide ``exact'' expectation values.
As for all the Monte Carlo methods, QMC results are affected by statistical uncertainties, 
but the property that makes some of them ``exact'' is the possibility of reducing 
within this unavoidable errors all the systematic errors introduced by the involved 
approximations.
This is true for Boson system at zero and at finite temperature, but the studies of Fermion 
systems, excited states or real time dynamics all suffer from sign problems which still have 
precluded the development of such kind of ``exact'' methods.
A first great subdivision of ``exact'' QMC methods is between finite and zero temperature 
methods.
Among $T=0$ K methods we find Green Function Monte Carlo (GFMC),\cite{gfmc1,gfmc2,gfmc3} 
Diffusion Monte Carlo (DMC),\cite{dmc} Path Integral Ground State Monte Carlo (PIGS)\cite{pigs} 
and Reptation Monte Carlo (RMC).\cite{rmc}
At finite temperature, the major role is played by Path Integral Monte Carlo (PIMC).\cite{pimc}

The main difference between finite and zero temperature methods is that ground state methods 
rely, to different extents, on a trial wave function for the importance sampling, while in PIMC 
no such wave function is involved at all.
PIMC needs only information on the interaction among particles as input, and the main 
difficulty to overcome, beyond the propagator accuracy, is the sampling ergodicity.
Recently a great step forward in this direction has been realized with the advent of the worm 
algorithm.\cite{worm}
The first GFMC computation did not use importance sampling,\cite{gfmc1} but this computation
was for a small system of 32 particles.
All the other computation at zero temperature have used a model wave function that was
given as input. 
With respect to PIGS method, if , in principle, convergence can be achieved for any
initial wave function that has a finite overlap with the ground state, the question is if
convergence can be achieved in a real computation.
Some progress have been achieved recently both for a finite\cite{cuer} and for a bulk Boson 
system.\cite{vita}  
For example, a Jastrow wave function, that describes the bulk phase of a quantum liquid, was 
employed as initial wave function in the study of small parahydrogen clusters.\cite{cuer} 
In the case of a two dimensional crystal of $^4$He, convergence to the same result was found
in the computation of the one--body density matrix starting from two initial wave functions 
with ``opposite'' properties,\cite{vita} one with Bose--Einstein condensate and the other
without.
Stimulated by these results, we have undergone a systematic check for a realistic Hamiltonian
for $^4$He and we have found that PIGS methods converge to the exact result regardless of the 
chosen initial wave function.  
Even better, PIGS converges even if the wave function contains wrong correlations or no 
correlations at all.

As a test--bed we have considered the bulk liquid and solid phases of a strongly interacting boson
system such as $^4$He.
By projecting very different wave functions, we find converging results for diagonal properties 
like the energy, the radial distribution function, the static structure factor and also for 
off-diagonal properties like the one-body density matrix.
For both the liquid and the solid phases we consider a shadow wave function (SWF),\cite{swf} which 
is the best available variational wave function\cite{moro} for $^4$He systems, and the constant 
wave function (CWF), which is the ground state wave function of the ideal Bose gas and does not 
contain any correlation.
The latter corresponds to have no importance sampling since the initial wave function gives the
same weight to any configuration of the particles.
Furthermore, for the liquid phase, we have considered also a localized wave function which 
describes an Einstein solid.
Our question is whether PIGS is able to recover the exact ground state starting from really 
different systems, i.e. an ideal gas or an Einstein crystal for the quantum liquid and an 
ideal gas for the quantum solid.
We find that for all the considered initial wave functions PIGS converges to the exact ground state.
Only the convergence rate depends on the initial wave function, and turns out to be slower for 
the worse ones.

The paper is organized as follows: Sec.~\ref{sec:meth} deals with the description of the PIGS 
method, of the test bed systems and of the initial wave functions.
Results are presented and discussed in Sec.~\ref{sec:results}.
Sec.~\ref{sec:conc} contains our conclusions.

\section{METHOD}
\label{sec:meth}

\subsection{The PIGS method}
\label{subsec:pigs}

It is well known that, given a Hamiltonian $\hat H$ for a quantum system of $N$ particles, the 
ground state wave function $\psi_0$ in the position representation can be obtained as the 
$\tau\to\infty$ limit of an imaginary time ($\tau=it/\hbar$) evolution of an initial (trial)
wave function $\psi_T$, provided that $\langle\psi_0|\psi_T\rangle\ne0$:
\begin{equation}
 \label{psi0}
  \psi_0(R) = \lim_{\tau\to\infty}\frac{e^{-\tau(\hat H-E_0)}\psi_T}{\langle\psi_0|\psi_T\rangle}
\end{equation}
Regardless of a normalization constant, which is not involved in the Monte Carlo sampling
and thus will be dropped in the following, an accurate approximation of the ground state wave 
function is given by
\begin{equation}
 \psi_\tau(R) = \int dR'\,G(R,R';\tau)\psi_T(R).
\end{equation}
This originates from the action of the imaginary time projector $\hat G=e^{-\tau\hat H}$, which 
exponentially removes from $\psi_T$ any overlap with the excited states during the imaginary time 
evolution. 
The evolved wave function $\psi_\tau$ and Eq. \eqref{psi0} provide the basis for all the zero 
temperature QMC methods.

A first problem rises: the imaginary time propagator $\hat G$ can be accurately written only for 
small values of $\tau$, but a large $\tau$ limit is necessary to ensure the convergence to 
$\psi_0$.
One possible strategy to overcome this problem is to reach the large $\tau$ limit by
means of a recursive procedure, as for example in GFMC\cite{gfmc1} and DMC.\cite{dmc}
Both these methods reach an extremely accurate approximation of the ground state wave 
function multiplied by $\psi_T$ by iterating equation \eqref{psi0} by means of random walks.
With these methods the trial wave function has a strategical role since it is involved at 
each iteration step, where it is used as importance sampling function.

On the contrary, the PIGS method\cite{pigs} interprets the imaginary time propagator as a density 
matrix operator corresponding to an inverse temperature $\beta=\tau$, and then, by exploiting 
the factorization property
\begin{equation}
 \label{factprop}
  G(R,R';\tau=\tau_1+\tau_2)=\int dR''\,G(R,R'';\tau_1)G(R'',R';\tau_2)
\end{equation}   
the large $\tau$ propagator $\hat G$ is written as a convolution of small imaginary time
propagators $G(R,R';\delta\tau)$ as in standard path integral formalism.\cite{pigs}
Thus the density matrix operator is broken up into $M$ small pieces with a time step 
$\delta\tau=\tau/M$ and the approximated ground state wave function reads
\begin{equation}
 \label{psitau}
  \begin{split}
   \psi_\tau(R) = & \int dR_1\cdots dR_M\, G(R,R_1;\delta\tau)G(R_1,R_2;\delta\tau)\cdots\\
                  & \times G(R_{M-1},R_M;\delta\tau)\psi_T(R_M).
 \end{split}
\end{equation}
An appealing feature, peculiar of this method is that, in $\psi_\tau$, the ansatz
on $\psi_T$ acts only at the starting point being the full imaginary time path 
governed by $\hat G$, which depends only on $\hat H$.
Once fixed $\delta\tau$, the elementary evolution step in imaginary time is obtained
acting with $G(R,R';\delta\tau)$ on a quantum state; this action is usually called projection step
due to the resulting increased overlap with the ground state. 
The PIGS method reaches convergence to the ground state when adding further imaginary time 
projections to $\psi_\tau$ does not provide any appreciable change of the results, i.e. the 
expectation values computed with $\psi_{\tau+ n \delta\tau}$ is compatible with those 
computed with $\psi_\tau$ within the statistical errors for any integer $n > 0$.
Since the convergence on the ground state for a finite system is exponentially fast,\cite{boni} 
the number $M$ of required imaginary time projections is usually limited.
Nevertheless $M$ depends on the specific expectation value one is computing, thus a separate analysis
of the convergence should be carried on for every computed quantity.
The convergence rate is determined essentially by the quality of the initial wave function.
The total imaginary time $\tau$ required to clean the initial wave function up from excited state 
contributions is smaller for good (i.e. with large overlap with the ground state) initial wave 
functions.
Also the accuracy of the approximation used for the small time $\hat G$ has a role in determining 
the computational cost of the projection procedure: in fact, with better approximations of the small 
imaginary time $\hat G$, one needs a smaller number of projection steps to reach convergence since
larger time step $\delta\tau$ are allowed.
Moreover the imaginary time projection procedure should be able, in principle, to incorporate 
in the final wave function all the correct correlations, even if these were absent in $\psi_T$.
Thus it should be possible, in principle, to obtain the correct ground state wave function, if
$\tau$ is large enough, even if $\psi_T$ is extremely poor, and the results do not suffer from
any variational bias. 
The intent of the present study is to show that this is indeed the case with values of $\tau$ 
manageable with present computational resources.

Like for finite temperature PIMC,\cite{pimc} also within PIGS the quantum system is mapped into 
a system of classical polymers, but in this case the polymers start and end on the initial wave 
function $\psi_T$,\cite{pigs,spig1,spig2} resulting in open linear polymers instead of the close 
ring ones of PIMC.\cite{pimc}
Each open polymer represents the full imaginary time path of a quantum particle that is sampled 
by means of the Metropolis algorithm.
Thus, the entire imaginary time evolution of the system is sampled at each Monte Carlo step, 
contrary to what happens for DMC, for example.
Expectation values of an operator $\hat O$ reads
\begin{equation}
 \label{aspett}
 \langle\hat O\rangle_\tau = \frac{\langle\psi_T|\hat G(\tau)\hat O\hat G(\tau)|\psi_T\rangle}
                             {\langle\psi_T|\hat G(\tau)\hat G(\tau)|\psi_T\rangle}
\end{equation}
and, if $\tau$ is large enough to allow convergence to the ground state, 
$\langle\hat O\rangle_\tau$ is the ground state expectation value, without needing
any extrapolation.
However an analysis of the convergence as a function of $\tau$ is needed, as 
explained before.

The ground state expectation value of the energy can be obtained in several ways:
the most largely employed one is the mixed estimate
\begin{equation}
 \label{mixed}
 \langle\hat H\rangle = \frac{\langle\psi_T|\hat G(2\tau)\hat H|\psi_T\rangle +
                              \langle\psi_T|\hat H\hat G(2\tau)|\psi_T\rangle}
                             {2\langle\psi_T|\hat G(2\tau)|\psi_T\rangle} .
\end{equation}
Notice that in PIGS this mixed estimate is exact.
In fact, since the Hamiltonian operator $\hat H$ commutes with the imaginary time evolution 
operator $\hat G$, it is possible to obtain unbiased expectation value of the Hamiltonian inserting
$\hat H$ at one of the ends of the path.
If $\psi_T$ is an accurate wave function, this estimator is preferable with respect to the 
direct one given by Eq.~\eqref{aspett} with $\hat O = \hat H$, because it has typically a 
lower variance.
This is no more true, however, when the initial wave function is particularly poor.
In this case the fluctuations in the expectation value of the Hamiltonian become so sizable
that \eqref{mixed} is unusable.
The direct estimator is seldom used because one has to compute the first and the second
derivative of $G(R,R';\delta\tau)$; but this is not a serious problem if the imaginary time 
propagator is available in an analytical formulation, like the one we employed here.
When $\psi_T$ is a good wave function, as for the SWF case, we have verified that mixed and 
direct estimators provide compatible results, and we report here the mixed one for the above 
mentioned reason.
On the contrary, for poor wave functions, like CWF, we had to consider the direct estimator.

As far as $^4$He systems are concerned, because of the Bose statistic obeyed by the atoms, 
one has, in principle, to account for permutations in the propagator $\hat G$.\cite{pigs,pimc}
Permutation moves are not strictly requested whenever the initial wave function has the 
correct Bose symmetry.
In fact, the polymer configuration resulting after a permutation can be in principle reached 
with a combination of standard sampling moves, since the polymers are open.\cite{spig1,spig2}
This is not the case if $\psi_T$ is not Bose symmetric like the Gaussian wave function: 
permutation cycles among particles must be introduced in the sampling in order to get the exact
ground state.
For all $\psi_T$ we have implemented permutation sampling following Ref.~\onlinecite{boni2}.
In fact, implementing the sampling of permutations even for a Bose symmetric $\psi_T$
turns out to greatly improve the ergodicity of the sampling.
In addition, we have used swap moves\cite{worm} because they increase the sampling efficiency
when computing off--diagonal properties.\cite{vita}

\subsection{Test systems}
\label{subsec:syst}

In order to test the PIGS method convergence properties we have considered two bulk phases of a
many--body strongly interacting Boson system: liquid and solid $^4$He.
Dealing with low temperature properties, $^4$He atoms are described as structureless zero--spin 
bosons, interacting through a realistic two--body potential, that we assume to be the HFDHE2 
Aziz potential~\cite{Aziz}.
For the liquid phase, we have considered a cubic box with periodic boundary conditions, 
containing $N=64$ atoms at the equilibrium density $\rho_l=0.0218$\AA$^{-3}$.
For the solid phase we have considered a cubic box with periodic boundary conditions designed to
house a fcc crystal of $N=32$ atoms at the density $\rho_s=0.0313$\AA$^{-3}$.
In both cases we add standard tail corrections to the potential energy to account for the finite 
size of the system by assuming the medium homogeneous (i.e. $g(r)=1$) beyond $L/2$, where $L$ is 
the size of the box.
Obviously, this is not an accurate assumption specially for the solid phase in such a small box, 
but our main purpose here is to show that PIGS method is able to reach the same results 
independently on the considered initial wave function.
Moreover we have studied the fcc lattice, which is stabilized by the cell geometry and by the 
periodic boundary conditions, whereas at zero temperature, the hcp lattice is the stable lattice.
Computations of ground state properties of bulk $^4$He with accurate tail corrections
can be found in the current literature.\cite{boro,moro}

\subsection{Initial wave functions}
\label{subsec:psit}

The standard initial wave functions commonly used within PIGS method\cite{pigs} are the  
variational Jastrow wave function (JWF) for the liquid and the Jastrow-Nosanow (J-NWF) for 
the solid.
A JWF represents the simplest possible choice of wave function for strongly interacting 
Bosons\cite{feen} and it contains only two--body correlations.
Using a McMillan pseudopotential,\cite{mcmil} the unnormalized JWF 
reads as
\begin{equation}
 \label{JWF}
 \psi_{\rm JWF}(R) = \prod_{i<j=1}^Ne^{-\frac{1}{2}\left(\frac{b}{r_{ij}}\right)^m}.
\end{equation}
The physical meaning of this JWF is that, due to the sharp repulsive part of the interaction
potential $V$ in the Hamiltonian $\hat H$, $^4$He atoms prefer to avoid each other.
In the J-NWF the JWF is multiplied by a term like the one in Eq.~\eqref{GWF} below, that
localizes the particles in a crystalline order.
In this work, however, in order to explore the convergence properties of the PIGS method, we
have considered two wave functions of ``opposite'' quality: the best available one, that is the 
shadow wave function, and the poorest imaginable one, i.e. the constant wave function.
JWF will be considered only when computing the one--body density matrix in the liquid phase.

The constant wave function is the ground state wave function of the ideal Bose gas, 
\begin{equation}
 \label{CWF}
 \psi_{\rm CWF}(R)=1.
\end{equation}
It carries no correlation at all.
We choose this wave function because, allowing an unrestricted sampling of the full 
configurational space, it results in no importance sampling.
Then the whole imaginary time projection procedure is driven only by the short time
evolution operator $\hat G$, without any input, and then any bias, from the initial state.
Thus at the starting point the system is made up by free particles; if after a long 
enough imaginary time projection, PIGS turns out to be able to reach a strong correlated 
quantum liquid and quantum crystal by itself we can safely believe that no variational bias 
affects PIGS results.

On the other hand, we choose as $\psi_T$ a SWF given by variational computation in order 
to have as reference results the ones coming from the projection of an initial wave function
that is more accurate as possible, i.e. from a wave function whose overlap with the exact 
ground state is known to be large.
In the SWF, additional correlations besides the standard two body terms are introduced via 
auxiliary variables which are integrated out.\cite{swf}
This is done so efficiently that the crystalline phase emerges as a spontaneously broken 
symmetry process, induced by the inter--particles correlations as the density is increased, 
without the need of any a priori knowledge of the equilibrium positions and without losing the 
translationally invariant form of the wave function.
Thus SWF is able to describe both the liquid and the solid phase with the same functional form
and it is explicitly Bose symmetric.
The standard SWF functional form reads
\begin{equation}
 \label{SWF}
 \psi_{\rm SWF}(R) = \phi_r(R)\int dS\,K(R,S)\phi_s(S)
\end{equation}
where $S=(\vec s_1,\vec s_2,\dots,\vec s_N)$ is the set of auxiliary shadow variables, 
$\phi_r(R)$ is the standard Jastrow two body correlation term \eqref{JWF}, $K(R,S)$ is a kernel 
coupling each shadow to the corresponding real variable, and $\psi_s(S)$ is another Jastrow term 
describing the inter--shadow correlations.
As usual,\cite{mcfar} we take $K(R,S)$ Gaussian and in $\phi_s(S)$ we use the rescaled and dilated 
He--He potential $V$ as pseudopotential.
The variational parameters we use were chosen in order to minimize the expectation value of the
Hamiltonian $\hat H$ and are reported in Ref.~\onlinecite{mcfar}.
Nowadays the SWF represents the best available variational wave function for $^4$He 
systems.\cite{moro} 
Recently, we have estimated\cite{vita} that, when describing a two dimensional solid, SWF 
overlap per particle with the true ground state is of about 99.8\%, which ensures a fast convergence 
rate when projected within the PIGS method.
The properties of the SWF are so peculiar that the PIGS method that has a SWF as $\psi_T$ 
deserves an its own name and is dubbed SPIGS: Shadow Path Integral Ground State 
method.\cite{spig1,spig2}
In the picture of linear polymers, the presence of the shadow variables adds two extra variational
links, one at each end of the polymer.

In order to test how robust PIGS is, we consider also a ``wrong'' wave function: for the liquid phase 
we consider a Gaussian wave function, where each particle is harmonically localized around  
fixed positions $\{\vec r_{0i}\}$
\begin{equation}
 \label{GWF}
 \psi_{\rm GWF}(R)=\prod_{i=1}^Ne^{-C|\vec r_i-\vec r_{0i}|^2},
\end{equation}
i.e. $\psi_T$ it the wave function of an Einstein harmonic solid.
The parameter $C=8$~\AA$^{-2}$ is arbitrary and it is was chosen to ensure a strong localization 
of the particles around the positions $\{\vec r_{0i}\}$ that were taken over a regular cubic lattice 
within the simulation box.
This wave function is evidently not translationally invariant and not Bose symmetric.
Furthermore it does not contain any correlation between the particles, and all the information 
that it carries is that of a crystalline system, i.e. GWF is an extremely poor wave function for 
the liquid phase.
This wrong initial wave function will provide a stringent test on the convergence properties of
the PIGS methods.

As far as the one--body density matrix computation in the liquid phase is concerned, the values 
of the parameters $b$ and $m$ in the JWF have been chosen equal to the ones of the corresponding
Jastrow term in the SWF.

\subsection{Small time propagator}
\label{subsec:G}

One of the fundamental elements of path integral projection Monte Carlo methods is the imaginary
time propagator $\hat G$, whose accuracy turns out to be crucial to the convergence to the exact 
results.
The functional form of $\hat G$ for a generic $\tau$ is unfortunately not known with exception of few 
particular cases, such as, for example, the free particle and the harmonic oscillator, but 
accurate approximations of $\hat G$ are obtainable in the small $\tau$ regime.\cite{pimc,boni,sakko}
In this work, we have chosen the Pair--Suzuki approximation\cite{pilat} for the imaginary time 
propagator, which is a pair--approximation of the fourth--order Suzuki--Chin density 
matrix.\cite{boni}

The Suzuki--Chin approximation is based on the following factorization of the density matrix:
\begin{equation}
 e^{-2\delta\tau\hat H}\simeq e^{-\frac{\delta\tau}{3}\hat V_e}
                       e^{-\delta\tau\hat T}
                       e^{-\frac{4\delta\tau}{3}\hat V_c}
                       e^{-\delta\tau\hat T}
                       e^{-\frac{\delta\tau}{3}\hat V_e}
\end{equation}
where $\hat T$ is the kinetic operator and $\hat V_e$ and $\hat V_c$ are given by
\begin{equation}
 \hat V_e=\hat V +\frac{\alpha\delta\tau^2\lambda}{3}\sum_{i=1}^N(\bf{F}_i)^2
\end{equation}
and
\begin{equation}
 \hat V_c=\hat V +\frac{(1-\alpha)\delta\tau^2\lambda}{6}\sum_{i=1}^N(\bf{F}_i)^2
\end{equation}
respectively, with $\hat V$ the potential operator, $\alpha$ an arbitrary constant in the 
range $[0,1]$, $\lambda=\hbar^2/2m$ and ${\bf F}_i={\bf \nabla}_i V$.
The resulting imaginary time propagator is accurate to order $\delta\tau^4$, and has been
successfully applied to liquid $^4$He in two and three dimensions.\cite{boni}
This approximation offers also the advantage that adjusting the parameter $\alpha$ it is possible
to optimize the convergence, and a standard choice for a quantum system is $\alpha=0$.\cite{boni}
A strategy to obtain a simpler, but equally accurate, approximation consists in applying a 
pair product assumption.\cite{pilat}
For sufficiently short time steps, in fact, the many--body propagator (in imaginary time) is well
approximated by the product of two--body propagators.\cite{pimc}
In this approximation, the small time propagator reads
\begin{equation}
 \label{pairS}
 \begin{split}
 G(R_m,R_{m+1};\delta\tau) =& \left(4\pi\lambda\delta\tau\right)^{-3N/2}\times \\
 &\prod_{i=1}^N \exp\left(-\frac{(\vec r_{i,m}-\vec r_{i,m+1})^2}{4\lambda\delta\tau}\right)\times \\
 &\exp\left(-u(r_{ij,m},r_{ij,m+1})\right)
 \end{split}
\end{equation}
where $u$ is given as
\begin{equation}
 \label{uPS}
 u(r_m,r_{m+1})=\left\{\begin{array}{ll}
 \frac{\delta\tau}{3}\left[v_e(r_{m})+2v_c(r_{m+1})\right] & m\quad {\rm odd}\\
 \\
 \frac{\delta\tau}{3}\left[2v_c(r_{m})+v_e(r_{m+1})\right] & m\quad {\rm even.}
 \end{array}\right.
\end{equation}
The potentials $v_e(r)$ and $v_c(r)$ are defined as
\begin{equation}
 \begin{split}
 & v_e(r) = V(r) + \alpha\frac{2}{3}\delta\tau^2\lambda\left(\frac{\partial V}{\partial r}\right)^2 \\
 & v_c(r) = V(r) + (1-\alpha)\frac{1}{3}\delta\tau^2\lambda\left(\frac{\partial V}{\partial r}\right)^2
 \end{split}
\end{equation}
where $V(r)$ is the potential experienced by two $^4$He atoms at a distance $r$.
The advantage is that there is no need to calculate $\bf{F}_i$.
As for the full Suzuki--Chin approximation,\cite{boni} also for the Pair--Suzuki the operators
corresponding to physical observables must be inserted only on odd time slices in the imaginary
time path.

\begin{figure}
 \includegraphics[width=8cm]{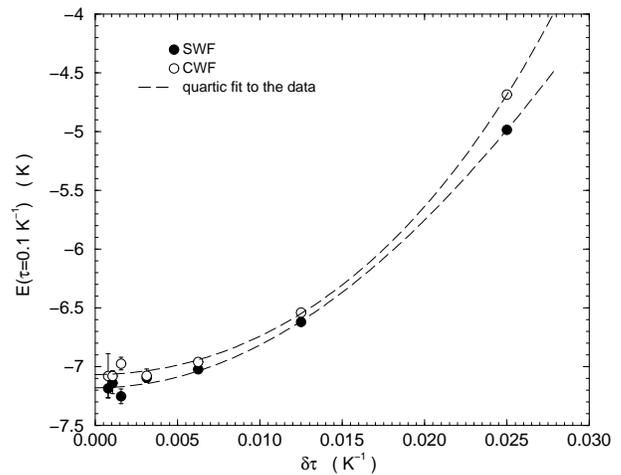}
 \caption{\label{f:dtau} Energy per $^4$He atom $E(\tau)$ vs. imaginary time step $\delta\tau$.
          The total projection time is $\tau=0.1$~K$^{-1}$.
          The calculations were carried out by projecting a SWF and a CWF for a system of 64
          particles at the equilibrium density $\rho=0.0218$~\AA$^{-3}$.
          Dashed lines are quartic fits to the data.
          Error bars, when not shown, are smaller than the used symbols.}
\end{figure}
In order to fix the optimal small imaginary time step value, we have performed PIGS simulations
with different initial wave functions.
By considering decreasing $\delta\tau$ values with a fixed total projection time, $\tau$, we 
have taken the energy per particle $E(\tau)$ as observable of reference.
As an example, our results for SWF and CWF in the liquid phase are plotted in Fig.~\ref{f:dtau}.
We choose as optimal value $\delta\tau=1/640$~K$^{-1}$; in fact, further reductions do not 
change the energy in a detectable way, i.e. within the statistical uncertainty.
In Fig.~\ref{f:dtau} SWF and CWF do not converge to the same value simply because the considered
total projection time $\tau$ in this test is not enough to ensure convergence of $E(\tau)$ to the 
ground state energy for CWF (see Fig.~\ref{f:enel}).
Similarly, in the solid phase we take $\delta\tau=1/960$~K$^{-1}$.

\section{RESULTS}
\label{sec:results}

Once set the optimal $\delta\tau$ value, we have computed the diagonal properties of the system 
for increasing total projection time $\tau$ until we reached convergence to a value that 
corresponds to the exact ground state result both for the liquid and for the solid phase.
In the liquid phase we have computed also the one--body density matrix.

\subsection{Liquid}
\label{sub:liq}

\subsubsection{PIGS results without importance sampling}

\begin{figure}
 \includegraphics[width=8cm]{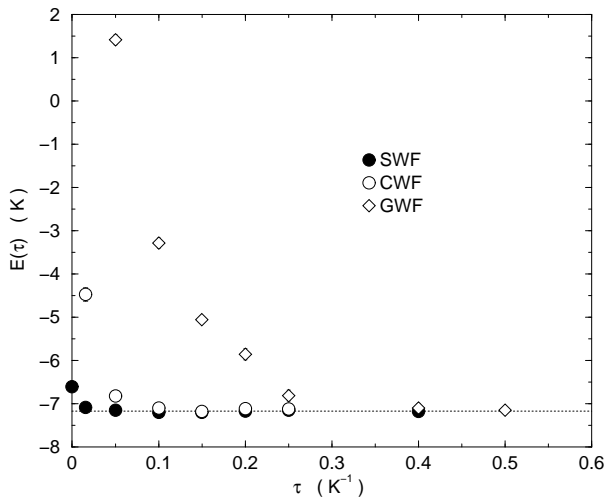}
 \caption{\label{f:enel} Energy per particle $E$ as a function of the total projection time 
          $\tau$ obtained from PIGS simulations for liquid $^4$He at the equilibrium density 
          $\rho=0.0218$~\AA$^{-3}$ by projecting a SWF (filled circles) and a CWF (open circles)
          and a GWF (open diamonds).
          $\tau=0$ result (filled circle) corresponds to the SWF variational estimate of $E$,
          the $\tau=0$ for the GWF is $E=122.08 \pm 0.06$ K and for CWF $E$ is essentially infinite.
          Error bars are smaller than the used symbols.
          Dotted line indicates the convergence value $E=-7.17\pm 0.02$ K.}
\end{figure}

For the liquid phase we have projected a SWF and a CWF.
The energy per particle as a function of the total projection time $\tau$ for both the wave 
functions is plotted in Fig.~\ref{f:enel}.
We find that the energy converges, independently from the considered initial wave function, to
the same value $E=-7.17\pm 0.02$~K.
This value, in spite of the small size of the considered system, is close to the 
experimental\cite{roach} result $E=-7.14$~K.
SWF converges very quickly, in fact $\tau=0.05$~K$^{-1}$ is already enough to ensure convergence.
CWF instead, requires a three times larger imaginary time, i.e. $\tau=0.15$~K$^{-1}$.
This was somehow expected, since SWF is presently the best available variational wave function for
$^4$He.\cite{moro}
Nevertheless, the quick convergence of also CWF is a really remarkable result.
In fact, this means that PIGS efficiently includes the exact interparticle correlations through
the imaginary time projections, without any need of importance sampling.
Then, the choice of a good wave function, within the PIGS method, becomes a matter of 
convenience rather than of principle, since better initial wave functions only allow for a 
smaller total projection time $\tau$, and thus less CPU consuming simulations.

\begin{figure}
 \includegraphics[width=8.2cm]{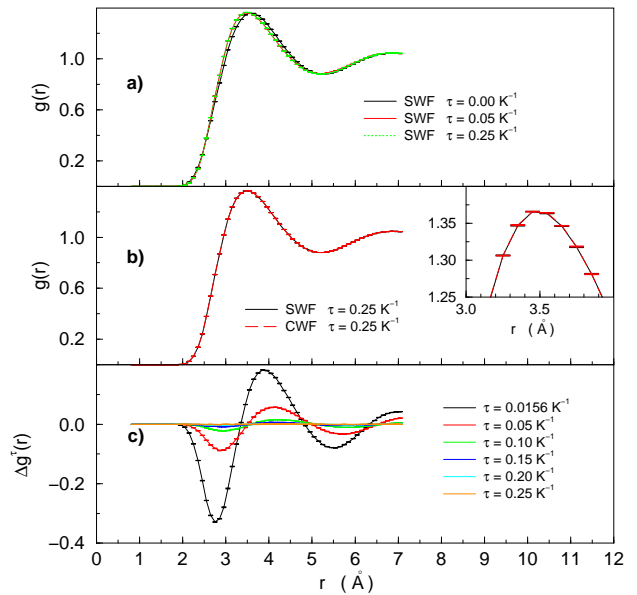}
 \caption{\label{f:grl} Radial distribution function $g(r)$ for bulk liquid $^4$He computed in 
          a cubic box with $N=64$ at the density $\rho=0.0218$~\AA$^{-3}$ with the PIGS method.
          a) $g(r)$ obtained by projecting a SWF for $\tau=0.00$, $0.05$ and $0.25$~K$^{-1}$.
          The $\tau=0.00$ result corresponds to the variational SWF estimate of $g(r)$.
          b) $g(r)$ obtained by projecting a SWF for $\tau=0.25$~K$^{-1}$ and a CWF for
          $\tau = 0.25$~K$^{-1}$.
          In the inset a zoom of the first maximum region.
          c) $\Delta g^\tau(r)=g_{\rm SWF}^\tau(r)-g_{\rm CWF}^\tau(r)$ at different $\tau$ 
          values, where $g_{\rm SWF}^\tau(r)$ is the $g(r)$ computed by projecting a SWF for an 
          imaginary time equal to $\tau$, and $g_{\rm CWF}^\tau(r)$ is the same but by 
          projecting a CWF.
          Note the smaller scale on the vertical axis}
\end{figure}
\begin{figure}
 \includegraphics[width=8.2cm]{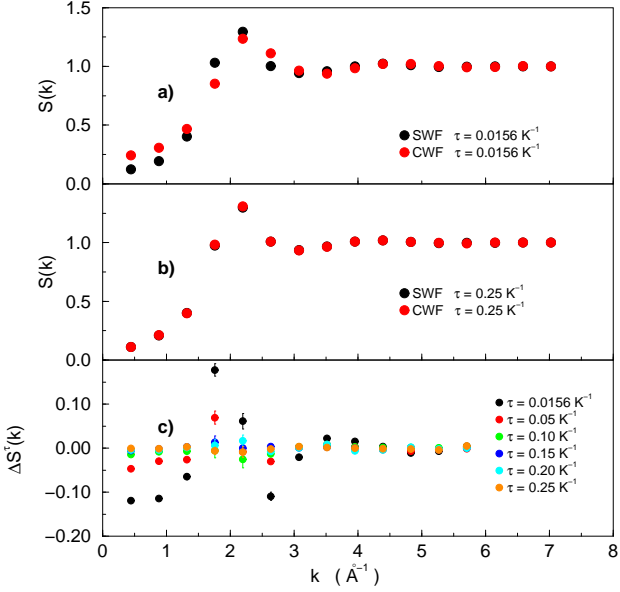}
 \caption{\label{f:skl} Static structure factor $S(k)$ for bulk liquid $^4$He computed in a
          cubic box with $N=64$ at the density $\rho=0.0218$~\AA$^{-3}$ with the PIGS method.
          a) $S(k)$ obtained by projecting a SWF and a CWF for $\tau=0.05$~K$^{-1}$. 
          b) $S(k)$ obtained by projecting a SWF and a CWF for $\tau=0.40$~K$^{-1}$.
          c) $\Delta S^\tau(k)=S_{\rm SWF}^\tau(k)-S_{\rm CWF}^\tau(k)$ at different $\tau$ values,
          where $S_{\rm SWF}^\tau(k)$ is the $S(k)$ computed by projecting a SWF for an
          imaginary time equal to $\tau$, and $S_{\rm CWF}^\tau(k)$ is the same but by
          projecting a CWF.
          Note the smaller scale on the vertical axis.}
\end{figure}

This convergence is confirmed also by the radial distribution function $g(r)$ and the static
structure factor $S(k)$.
For such quantities, the convergence rate is found to be similar to the energy one.
In Fig.~\ref{f:grl} we report the radial distribution function $g(r)$ obtained by projecting 
both a SWF and a CWF at different imaginary time values.
For $\tau>0.05$~K$^{-1}$, SWF results at different $\tau$ are indistinguishable within 
the statistical uncertainty (see Fig.~\ref{f:grl}a).
In fact, with SWF the exact result is reached within very few projection steps and then it is 
no more affected by further projections.
As already pointed out, also CWF displays a fast convergence, as shown in Fig.~\ref{f:grl}c, 
where $\Delta g^\tau(r)=g_{\rm SWF}^\tau(r)-g_{\rm CWF}^\tau(r)$ is shown.
For increasing $\tau$, $\Delta g^\tau$ evolves toward a flat function, meaning that the systems 
described starting from the two different wave functions, i.e the strongly correlated quantum 
liquid of SWF and the ideal gas of CWF, are evolving into the same quantum liquid, which is the
best reachable representation of the exact ground state of the simulated system.
The same conclusion is inferred from the evolution of the static structure factor $S(k)$, 
which is plotted in Fig.~\ref{f:skl}.

\subsubsection{PIGS results from a wrong initial function}

In order to put a more stringent check on the PIGS method ability to converge to
the exact ground state without any variational bias, we have considered also a wrong initial
wave function by projecting a GWF.
Thus at the starting point of the imaginary time path there is now a strongly localized 
Einstein crystal.
We find, even in this case, that the energy converges to the same value as before (see 
Fig.~\ref{f:enel}).
Thus PIGS is able not only to drop from the initial wave function the wrong information of
localization, but also to generate at the same time the correct correlations among the particles.
GWF needs $\tau=0.5$~K$^{-1}$ to converge, which is ten times larger than the SWF value.

\begin{figure}
 \includegraphics[width=8.2cm]{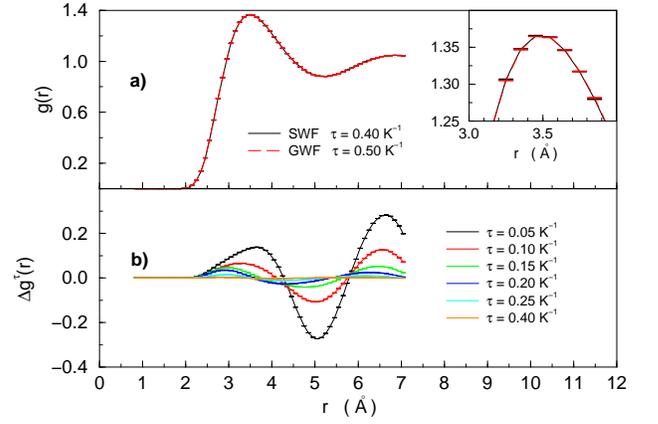}
 \caption{\label{f:grl2} Radial distribution function $g(r)$ for bulk liquid $^4$He computed in 
          a cubic box with $N=64$ at the density $\rho=0.0218$~\AA$^{-3}$ with the PIGS method.
          a) $g(r)$ obtained by projecting a SWF for $\tau=0.40$~K$^{-1}$ and a GWF for
          $\tau = 0.50$~K$^{-1}$.
          In the inset a zoom of the first maximum region.
          b) $\Delta g^\tau(r)=g_{\rm SWF}^\tau(r)-g_{\rm GWF}^\tau(r)$ at different $\tau$ values,
          where $g_{\rm SWF}^\tau(r)$ is the $g(r)$ computed by projecting a SWF for an 
          imaginary time equal to $\tau$, and $g_{\rm GWF}^\tau(r)$ is the same but by 
          projecting a GWF.
          Note the smaller scale on the vertical axis.}
\end{figure}
\begin{figure}
 \includegraphics[width=8.2cm]{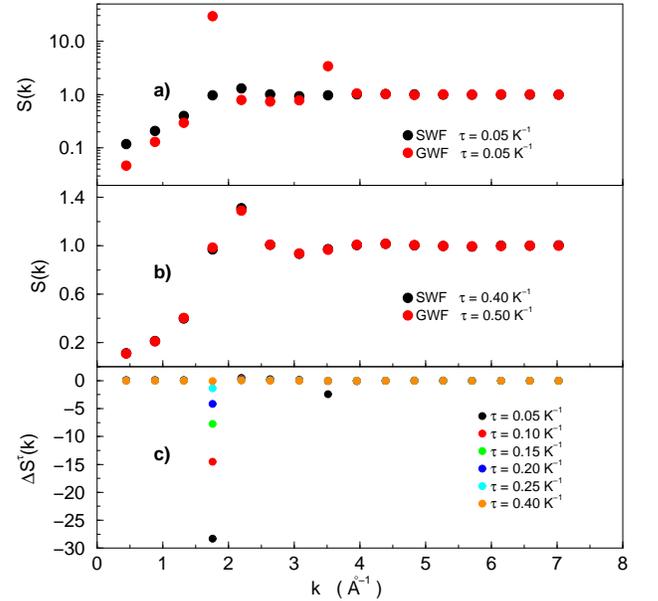}
 \caption{\label{f:skl2} Static structure factor $S(k)$ for bulk liquid $^4$He computed in a
          cubic box with $N=64$ at the density $\rho=0.0218$~\AA$^{-3}$ with the PIGS method.
          a) $S(k)$ obtained by projecting a SWF and a GWF for $\tau=0.05$~K$^{-1}$. 
          It is evident in the GWF result the presence of the Bragg peak.
          Note the logarithmic scale.
          b) $S(k)$ obtained by projecting a SWF for $\tau=0.40$~K$^{-1}$ and a GWF for 
          $\tau=0.50$~K$^{-1}$.
          The Bragg peak is no more present in the GWF result.
          c) $\Delta S^\tau(k)=S_{\rm SWF}^\tau(k)-S_{\rm GWF}^\tau(k)$ at different $\tau$ values,
          where $S_{\rm SWF}^\tau(k)$ is the $S(k)$ computed by projecting a SWF for an
          imaginary time equal to $\tau$, and $S_{\rm GWF}^\tau(k)$ is the same but by
          projecting a GWF. 
          Note the change of the vertical scale.
          Error bars are smaller than the used symbols.}
\end{figure}

Again this convergence is confirmed also by the radial distribution function $g(r)$ and the static
structure factor $S(k)$.
In Fig.~\ref{f:grl2} we report the radial distribution function $g(r)$ obtained by projecting 
a GWF at different imaginary time values compared with the ones coming from the projection of SWF.
It is evident that small imaginary time is not enough to leave out the wrong information 
in the GWF.
For lower $\tau$ values, there are still reminiscences of the starting harmonic solid, which 
are progressively lost as the projection time increases.
This is made clearer in Fig.~\ref{f:grl2}b where we plot the difference $\Delta g^\tau(r)$, 
at fixed imaginary time $\tau$, between the $g(r)$ computed by projecting the SWF and the one 
obtained by projecting the GWF.
A similar behavior is observed in the evolution static structure factor $S(k)$, plotted in 
Fig.~\ref{f:skl2}.
For the GWF, the Bragg peak shown at small $\tau$ values (Fig.~\ref{f:skl2}a), which is typical 
of the solid phase, becomes lower and lower as the projection time is increased 
(Fig.~\ref{f:skl2}b), until convergence is reached (see Fig.~\ref{f:skl2}c).

From the plot of the energy per particle vs. the total imaginary time $\tau$ it is possible to 
estimate the overlap per particle of the initial wave function on the exact ground state.\cite{mora}
By using the results in Fig.~\ref{f:enel} we find that the overlap of SWF is about 99\%, while
the GWF one is about 10\%.
That SWF has an high overlap with the ground state is not a surprise; it was qualitatively 
expected since SWF is presently the best available wave function for $^4$He.\cite{moro}
However a 99\% overlap is really remarkable and provides a further argument on the goodness of
SWF.
On the other hand, a poor overlap of GWF was somehow expected, since the parameter $C$ was 
chosen to strongly localize the atoms of the bulk liquid around fictitious equilibrium 
positions on a regular lattice.

\subsubsection{Off-diagonal properties}

\begin{figure}
 \includegraphics[width=8cm]{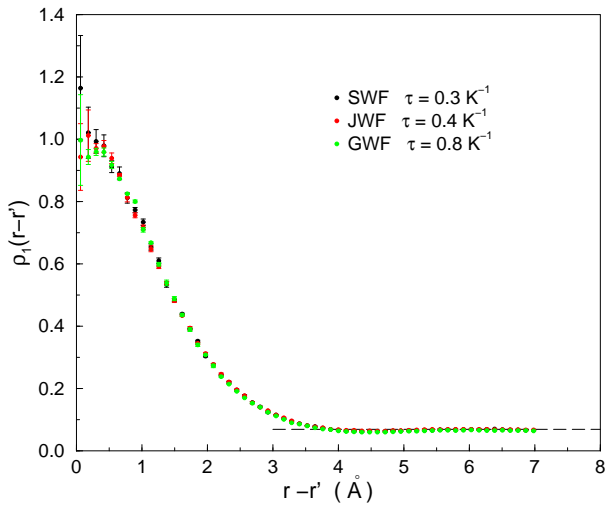}
 \caption{\label{f:odlro} One--body density matrix $\rho_1$ obtained from PIGS simulations for 
          liquid $^4$He at the equilibrium density $\rho=0.0218$~\AA$^{-3}$ by projecting a SWF,
          a JWF and a GWF for an imaginary time $\tau=0.30$, $0.40$ and $0.80$~K$^{-1}$ 
          respectively.
          The dotted line indicates the condensate value $n_0=0.069$ obtained from an independent
          PIGS simulation.\cite{moro2}}
\end{figure}
Besides the diagonal ones, also off--diagonal properties, such as the one--body density matrix,
are accessible within PIGS simulations.
The one-body density matrix $\rho_1(\vec r,\vec r')$ represents the probability amplitude of
destroying a particle in $\vec r$ and creating one in $\vec r'$.
Its Fourier transformation represents the momentum distribution.
In first quantization $\rho_1$ is given by the overlap between the normalized many-body ground
state wave functions $\psi_0(R)$ and $\psi_0(R')$, where the configuration
$R'=(\vec r',\vec r_2,\dots,\vec r_N)$ differs from  $R=(\vec r,\vec r_2,\dots,\vec r_N)$ only
by the position of one of the $N$ atoms in the system.
If $\psi_0(R)$ is translationally invariant, $\rho_1$ only depends on the difference
$|\vec r-\vec r'|$, thus
\begin{equation}
 \label{eq:obdm}
  \rho_1(\vec r-\vec r')=N\int d\vec r_2\dots d\vec r_N\,\psi_0^*(R)\psi_0(R').
\end{equation}
The Bose-Einstein condensate fraction $n_0$ is equal to the large distance limit of
$\rho_1(\vec r-\vec r')$.
In fact, if $\rho_1$ has a nonzero plateau at large distance, the so called
off-diagonal long-range order (ODLRO), its FT contains a Dirac delta function,
which indicates a macroscopic occupation of a single momentum state, i.e. Bose--Einstein
condensation.

The exact $\rho_1$ can be obtained in PIGS simulation by substituting $\psi_0$ in
\eqref{eq:obdm} with $\psi_\tau$ with $\tau$ large enough.
This corresponds to the simulation of a system of $N-1$ linear polymers plus a polymer which is
cut into two halfs, called half--polymers, one departing from $\vec r$ and the other from
$\vec r'$.
Thus $\rho_1$ is obtained by collecting the relative distances among the cut ends of the two 
half--polymers during the Monte Carlo sampling.
The present computation of $\rho_1$ has been obtained by implementing a zero temperature version
of the worm algorithm.\cite{worm}
We have worked with a fixed number of particles and not in the grand canonical ensemble, similarly
to what has been done at finite temperature in Ref.~\onlinecite{pilat}.
In practice this corresponds to a usual PIGS calculation of $\rho_1$ where ``open'' and ``close''
moves have been implemented\cite{worm} in order to visit diagonal and off-diagonal sectors within
the same simulation. 
The advantage of doing this does not come from the efficiency of the worm algorithm to explore 
off-diagonal configurations, because similar efficiency is obtained with PIGS when ``swap'' moves 
are implemented.\cite{vita} 
The benefit in using a worm-like algorithm here instead comes from the automatic normalization of 
$\rho_1$ which is a peculiarity of this method.\cite{worm}
In Fig.~\ref{f:odlro} we report $\rho_1$ obtained in PIGS simulations of bulk liquid $^4$He
at $\rho=0.0218$~\AA$^{-3}$ by projecting either a SWF, a JWF and a GWF.
All the simulations give the same result, shown in Fig.~\ref{f:odlro} which turns out to be
compatible with the recent estimate obtained with PIGS given in Ref.~\onlinecite{moro2} of 
$n_0=0.069 \pm 0.005$.

\subsection{Solid}
\label{sub:sol}

\begin{figure}
 \includegraphics[width=8cm]{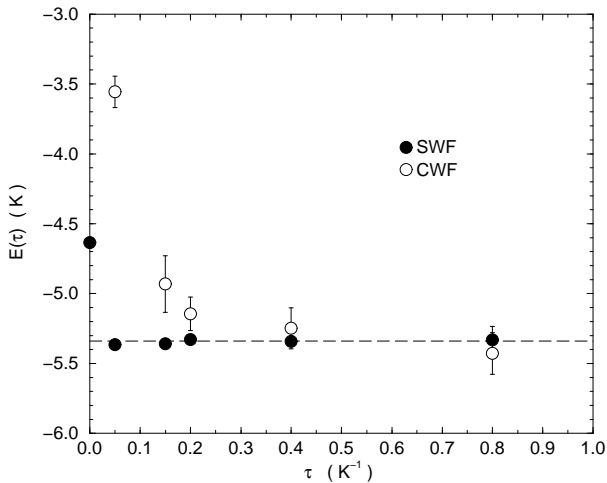}
 \caption{\label{f:enes} Energy per particle $E$ as a function of the total projection time
          $\tau$ obtained from PIGS simulations of an fcc $^4$He crystal at the density
          $\rho=0.0313$~\AA$^{-3}$ by projecting a SWF (filled circles) and a CWF (open circles).
          Dashed line indicates the convergence value $E=-5.34\pm 0.02$~K.}
\end{figure}

\begin{figure}
 \includegraphics[width=8.2cm]{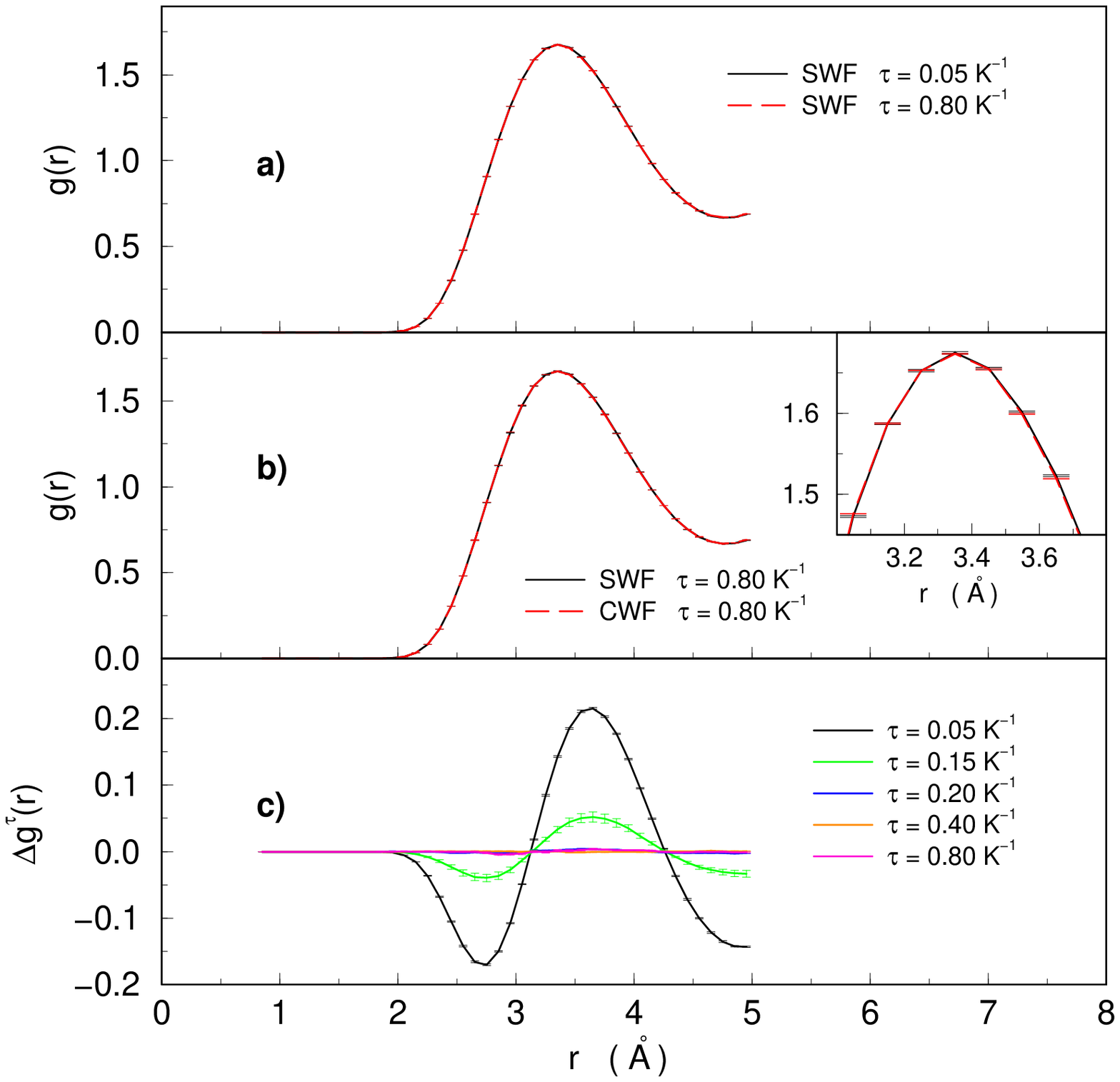}
 \caption{\label{f:grs} Radial distribution function $g(r)$ for bulk solid $^4$He computed in a
          cubic box with $N=32$ at the density $\rho=0.0313$~\AA$^{-3}$ with the PIGS method.
          a) $g(r)$ obtained by projecting a SWF for $\tau=0.05$ and $0.80$~K$^{-1}$.
          b) $g(r)$ obtained by projecting a SWF and a CWF for $\tau=0.80$~K$^{-1}$.
          In the inset a zoom of the first maximum region.
          c) $\Delta g^\tau(r)=g_{\rm SWF}^\tau(r)-g_{\rm CWF}^\tau(r)$ at different $\tau$ values,
          where $g_{\rm SWF}^\tau(r)$ is the $g(r)$ computed by projecting a SWF for an imaginary
          time equal to $\tau$, and $g_{\rm CWF}^\tau(r)$ is the same but by projecting a CWF.}
\end{figure}

\begin{figure}
 \includegraphics[width=8.2cm]{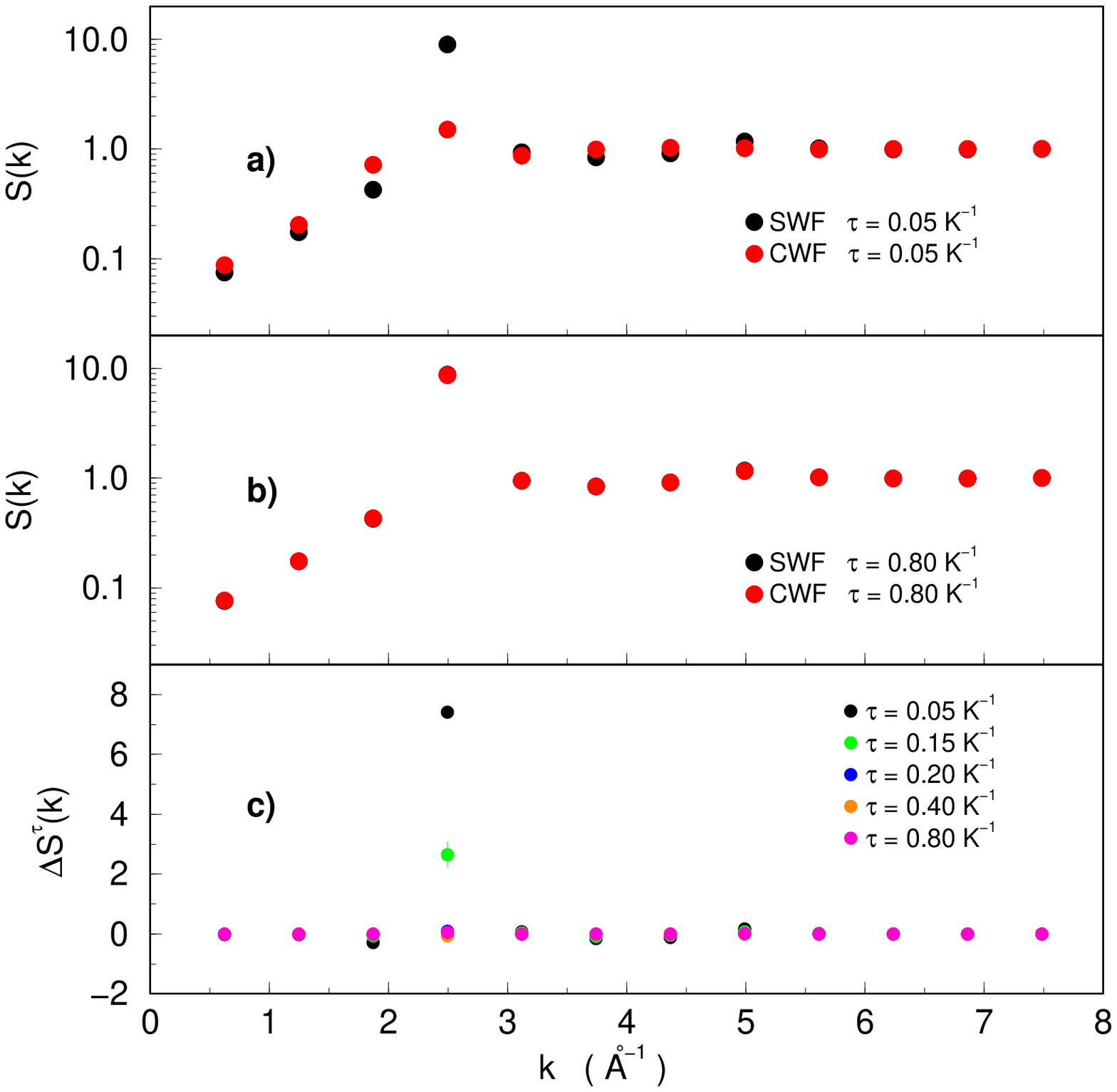}
 \caption{\label{f:sks} Static structure factor $S(k)$ for bulk solid $^4$He computed in a
          cubic box with $N=32$ at the density $\rho=0.0313$~\AA$^{-3}$ with the PIGS method.
          a) $S(k)$ obtained by projecting a SWF and a CWF for $\tau=0.05$~K$^{-1}$.
          b) $S(k)$ obtained by projecting a SWF and a CWF for $\tau=0.80$~K$^{-1}$. The black
             dots are under the red ones.
          c) $\Delta S^\tau(k)=S_{\rm SWF}^\tau(k)-S_{\rm CWF}^\tau(k)$ at different $\tau$ values,
          where $S_{\rm SWF}^\tau(k)$ is the $S(k)$ computed by projecting a SWF for an imaginary
          time equal to $\tau$, and $S_{\rm CWF}^\tau(k)$ is the same but by projecting a CWF.
          Error bars are smaller than the used symbols.
          Notice the logarithmic scale in panels a) and b).}
\end{figure}

We have performed the computation at density $\rho=0.0313$~\AA$^{-3}$, where $^4$He is in the 
solid phase, by projecting a SWF and a CWF.
Our results for the energy per particle are plotted in Fig.~\ref{f:enes} as a function of $\tau$.
In both cases we find convergence to the value $E=-5.34\pm 0.02$~K. 
Even in this phase the convergence of SWF is faster, being $\tau=0.05$~K$^{-1}$ enough to reach 
convergence.
In the case of CWF convergence is reached only for a much larger imaginary time 
$\tau=0.80$~K$^{-1}$.

Also in this case convergence is obtained for the radial distribution function and for the 
static structure factor, reported in Fig.~\ref{f:grs} and Fig.~\ref{f:sks} respectively.
From Fig.\ref{f:grs}a it is evident that SWF has reached the true ground state with few projection 
steps, since the results for $g(r)$ at $\tau=0.05$~K$^{-1}$ and $\tau=0.80$~K$^{-1}$ are 
indistinguishable.
The evolution toward the correct ground state of the projected CWF is instead detectable.
The presence of the crystalline structure is mainly evident in the static structure factor, 
where a Bragg peak grows with increasing $\tau$ (see Fig.~\ref{f:sks}a,b).
The emerging of the correct solid structure by projecting a really poor wave function such as the 
CWF is made evident by the trend toward a flat function of the differences $\Delta g^\tau(r)$ and
$\Delta S^\tau(k)$ plotted in Fig.\ref{f:grs}c and Fig.\ref{f:sks}c respectively.

\section{CONCLUSION} 
 \label{sec:conc}

In this work we have studied with the Path Integral Ground State method
diagonal and off-diagonal properties of a strongly interacting
quantum Bose system like the bulk liquid and solid phases of $^4$He.
We have obtained convergence to the ground state values of quantities like the total energy,
the radial distribution function, the static structure factor and the one-body density matrix
projecting radically different wave functions: equivalent expectation values in the liquid phase 
have been obtained using as initial wave function a shadow wave function, a Gaussian wave function 
with strongly localized particles of an Einstein solid without interparticle correlations and
also a constant wave function where all configurations of the particles are equally probable.
Similarly in the solid phase equivalent expectation values have been obtained by considering a 
shadow wave function, which describes a solid, and a constant wave function which describes
an ideal Bose gas.
The present analysis demonstrates the absence of any variational bias in PIGS; a method that
can be thus considered as unbiased as the finite temperature PIMC.
This remarkable property comes from the accurate imaginary time propagators, exactly the same
used with PIMC, that do not depend on the initial trial state.
It remains true that the use of a good variational initial wave function greatly improves the
rate of convergence to the exact results.
Moreover, very poor wave functions have also the drawback of requiring the direct estimator for 
the Hamiltonian (Eq.~(\ref{aspett}) with $\hat O=\hat H$) implying the necessity of an analytical 
formulation for the small imaginary time propagator $\hat G$ or an accurate knowledge of its 
derivatives.

We have addressed here only the case of a realistic interaction potential among Helium atoms. 
However one can reasonably expect that this conclusion holds even for very different kinds of 
interaction, once an accurate approximation for the imaginary time propagator is known (for 
example hard-spheres\cite{hard} or hydrogen plasma\cite{pier}).
As far as pathological potentials like the attractive Coulomb one are concerned, PIGS would suffer
the same limitations of PIMC if inaccurate approximations of the propagator were used.\cite{coul}

\section{ACKNOWLEDGMENTS}
 \label{sec:akcno}

Authors acknowledge S. Pilati for useful discussions.
This work was supported by the INFM Parallel Computing Initiative, by the 
Supercomputing facilities of CILEA and by the Mathematics Department
``F. Enriques'' of the Universit\`a degli Studi di Milano.

\end{document}